\documentclass{article}

\usepackage{microtype}
\usepackage{nicefrac}
\usepackage{enumitem}
\usepackage{graphicx}
\usepackage{amsmath}
\usepackage{subfigure}
\usepackage{booktabs} % for professional tables
\usepackage{array}
\usepackage{caption}
\usepackage[title, titletoc]{appendix}
\newcolumntype{L}{>{\centering\arraybackslash}m{2cm}}
\usepackage{url}

\usepackage{breakurl}
\usepackage[breaklinks]{hyperref}

\usepackage[accepted]{icml2021}

\icmltitlerunning{Towards Quantifying the Carbon Emissions of Differentially Private Machine Learning}%Understanding the impacts of Differentially Private Networks on the Carbon Footprint}

\begin{document}

\twocolumn[
\icmltitle{%Understanding the impacts of Differentially Private Networks on\\ the Carbon Footprint}
Towards Quantifying the Carbon Emissions of\\Differentially Private Machine Learning}

% It is OKAY to include author information, even for blind
% submissions: the style file will automatically remove it for you
% unless you've provided the [accepted] option to the icml2021
% package.

% List of affiliations: The first argument should be a (short)
% identifier you will use later to specify author affiliations
% Academic affiliations should list Department, University, City, Region, Country
% Industry affiliations should list Company, City, Region, Country

% You can specify symbols, otherwise they are numbered in order.
% Ideally, you should not use this facility. Affiliations will be numbered
% in order of appearance and this is the preferred way.
\icmlsetsymbol{equal}{*}

\begin{icmlauthorlist}
\icmlauthor{Rakshit Naidu}{equal,cmu,man}
\icmlauthor{Harshita Diddee}{equal,bhar}
\icmlauthor{Ajinkya Mulay}{equal,purd}\\
\icmlauthor{Aleti Vardhan}{equal,man}
\icmlauthor{Krithika Ramesh}{equal,man}
\icmlauthor{Ahmed Zamzam}{nrel}
% \icmlauthor{Tateu H.~Yasehe}{ed,to,goo}
% \icmlauthor{Aaoeu Iasoh}{goo}
% \icmlauthor{Buiui Eueu}{ed}
% \icmlauthor{Aeuia Zzzz}{ed}
% \icmlauthor{Bieea C.~Yyyy}{to,goo}
% \icmlauthor{Teoau Xxxx}{ed}
% \icmlauthor{Eee Pppp}{ed}
\end{icmlauthorlist}

\icmlaffiliation{cmu}{Carnegie Mellon University}
\icmlaffiliation{bhar}{Bharati Vidyapeeth's College of Engineering}
\icmlaffiliation{man}{Manipal Institute of Technology}
\icmlaffiliation{purd}{Purdue University}
\icmlaffiliation{nrel}{The National Renewable Energy Laboratory}

\icmlcorrespondingauthor{Rakshit Naidu}{rnemakal@andrew.cmu.edu}
\icmlcorrespondingauthor{Harshita Diddee}{harshitadd@gmail.com}
\icmlcorrespondingauthor{Ajinkya Mulay}{mulay@purdue.edu}

% You may provide any keywords that you
% find helpful for describing your paper; these are used to populate
% the "keywords" metadata in the PDF but will not be shown in the document
\icmlkeywords{Machine Learning, ICML}

\vskip 0.3in
]

% this must go after the closing bracket ] following \twocolumn[ ...

% This command actually creates the footnote in the first column
% listing the affiliations and the copyright notice.
% The command takes one argument, which is text to display at the start of the footnote.
% The \icmlEqualContribution command is standard text for equal contribution.
% Remove it (just {}) if you do not need this facility.

%\printAffiliationsAndNotice{}  % leave blank if no need to mention equal contribution
\printAffiliationsAndNotice{\icmlEqualContribution} % otherwise use the standard text.

\begin{abstract}
In recent years, machine learning techniques utilizing large scale datasets have achieved remarkable performance. Differential privacy, by means of adding noise, provides strong privacy guarantees for such learning algorithms. The cost of differential privacy is often a reduced model accuracy and a lowered convergence speed. This paper investigates the impact of differential privacy on learning algorithms in terms of their carbon footprint due to either longer run-times or failed experiments. Through extensive experiments, further guidance is provided on choosing the noise levels which can strike a balance between desired privacy levels and reduced carbon emissions.
\end{abstract}

\vspace{-9mm}

\section{Introduction}
With the rising availability of large-scale, diverse datasets, performance of Machine Learning (ML) models have experienced a significant boost across a multitude of domains. This boost is also associated with the availability of extreme-scale datasets, which is heavily linked to individual user contributions achieved via crowd-sourcing. ML algorithms often perform operations directly on raw user data leading to a host of privacy violations. Differential Privacy (DP) \cite{DworkRoth14, Abadi_2016} makes progress in this domain by providing strong privacy guarantees for such contributing individuals. This guarantee is achieved by means of noise addition, which can be done at various stages of the ML pipeline including : (1) \textit{Local DP:} Addition to the raw data \cite{cormode2018privacy} (2) \textit{Gradient DP:} Addition  to gradients after clipping \cite{Abadi_2016} (3) \textit{ Addition to Output \& Objective DP:} Addition to the final ML model or the loss function \cite{chaudhuri2011differentially}.

\subsection{Impact on Climate Change }

It is well-known that the computational resource investment requisite for training ML models generates a carbon footprint. This footprint is amplified in privacy-preserving setups where it is harder to reach consistent accuracy due to the addition of noise. Extended and failed runs (especially on larger datasets) actively contribute to an increase in the carbon footprint of ML experiments \cite{EnergyNLP19}. Therefore, an analysis of the climatic impact of this privacy modulation is critical. While the existing DP literature studies several performance aspects affected by varying privacy requirements, it lacks a comprehensive quantification of the carbon footprint of DP and how it is affected by variable privacy levels. Since DP also provides a mathematical paradigm to quantify the privacy budget of training ML models while tracking the privacy usage across multiple runs, this paper aims at quantifying the Carbon Emissions (CE) associated with varying privacy budgets of differentially private networks. In order to study impact of DP on these emissions, we implement \textit{Gradient DP} (DP-SGD \cite{Abadi_2016}) for natural language processing, image classification, and reinforcement learning domains to identify the privacy implications, model performance and most crucially the carbon footprint of each algorithm. As per our knowledge this is the first attempt to quantitatively benchmark the carbon footprint of differentially private ML models.

\subsection{Differential Privacy}
\textit{Definition 1}: Given a randomized mechanism $\mathcal{A}: \mathcal{D} \rightarrow \mathcal{R}$ (with domain $\mathcal{D}$ and range $\mathcal{R}$) and any two neighboring datasets $d_{1}, d_{2} \in \mathcal{D}$ (\emph{i.e.} they differ by a single individual data element), $\mathcal{A}$ is said to be $(\varepsilon, \delta)$-differentially private for any subset $S \subseteq \mathcal{R}$ \footnote{In this work, we exclusively use Gaussian noise}.
\begin{equation}
\Pr\left[ \mathcal{A}\left( d_{1}\right) \in S\right] \leq e^{\varepsilon }\cdot\Pr\left[ \mathcal{A}\left( d_{2}\right) \in S\right] + \delta
\end{equation}
Here, $\epsilon \geq 0, \delta \geq 0$. A $\delta = 0$ case corresponds to pure differential privacy, while both $\epsilon = 0, \delta = 0$ leads to an infinitely high privacy domain. Finally, $\epsilon = \infty$ provides no privacy guarantees. For practical purposes we want $\epsilon \leq 5, \delta \ll \frac{1}{N}$ where $N$ is the number of samples in the dataset. \cite{DworkRoth14}

The privacy of differentially private models can be quantified with parameters such as epsilon ($\varepsilon$) and delta ($\delta$). Utilizing DP-SGD \cite{Abadi_2016}, that is, adding noise to the gradients at each step during training using a clipping factor ($S$) and noise multiplier ($z$), the amount of noise added to the model can be linked to to the degree of privacy that the model can achieve. Theoretically, a lower value of $\varepsilon$ indicates a higher degree of privacy and this increased privacy degree is understandably, achieved at the expense of model performance due to the addition of the noise. The practical implication of this, however, includes a direct impact on the computational resources required to achieve model performance. Reduced privacy requirements allow the addition of noise with limited power, and hence, models can achieve appropriate performance without any significant resource expense. On the other hand, high privacy requirements necessitate adding a significantly large magnitude of noise which may directly lead to an increase in the number of training passes that the model has to iterate over to achieve the same accuracy. Further, noise addition may even lead to the non-convergence of some systems in the worst case.

\subsection{Related Work}

Works such as \cite{EnergyNLP19, CO2Forbes20} discuss how conventional Machine Learning models impact carbon footprint. 
In particular, \cite{EnergyNLP19} discusses how training a single Deep Learning model generates the total lifetime carbon footprint of nearly five cars (as mentioned in \cite{CO2Forbes20}) which is more than 17 times the amount of CO$_2$ emissions generated by an average American per year.
Regarding DP, there has been very little considerations on how Privacy-Preserving Machine Learning (PPML) impacts climate change. In \cite{FLClimateChange20}, a comprehensive study is presented on how local client-side models in Federated learning (FL) could potentially hold quality data required to understand climate change given data privacy concerns due to recent policies like GDPR \cite{GDPR18}. However, running local models on multiple client devices and aggregating them globally at the server level requires additional infrastructure in place, thereby causing a detrimental effect on carbon emissions.

\subsection{Contributions and Impacts}
In this paper, we provide the first benchmark to quantitatively assess how DP-noise affect carbon emissions in three different tasks : (1) a Natural Language Processing (NLP) task using news classification (2) a Computer Vision (CV) task using the MNIST dataset and (3) a Reinforcement Learning (RL) task using the Cartpole control problem. Intuitively, when DP noise is added to ML pipelines, the carbon emissions should increase as the energy required for computations increase due to rising number of epochs required for convergence. In order to quantify how the addition of noise plays into climate change, we track carbon emissions in the models using the \textit{codecarbon} tool \cite{codecarbon}, a joint effort from authors of \cite{lacoste2019quantifying} and \cite{lottick2019nergy}. We record the average accuracy of several runs of the considered ML task to assess the behavior of DP-noise.

Noise for masking data has been widely used in adversarial machine learning \cite{kurakin2016adversarial}. Given the rise in Privacy-enhancing Technologies and privacy policies, noise addition has now become prevalent in the context of DP. We envisage this work to provide an insight on how much noise could result in varying amounts of CO$_2$ emissions. Hence, our work takes a peek at how the addition of noise could impact a number of industries from healthcare to finance and justice, where sensitive data is commonly in use.

\section{Experimental Results}

\subsection{BERT}

In these set of experiments, we evaluate the performance of two experiments on Bidirectional Encoder Representations from Transformers or BERT \cite{devlin-etal-2019-bert}. The model is fine-tuned for topic-classification of news articles. The primary objective of these experiments is to observe the carbon emissions and energy usage of vanilla BERT and DP-BERT (over different privacy levels).
\par A randomly down-sampled subset (15,000 samples) of the AG News Classification \cite{agnewsclass} is used for this task with a 80/20 train-test split. We use BERT with the AdamW optimizer with the \textit{bert-base-cased} tokenizer (batch size ($B$) of $32$, Learning Rate being  0.0005) to conduct the following experiments for this task.

\subsubsection{Carbon emissions and energy consumed}
    The aim of this experiment is to analyse any possible association between different levels of privacy and carbon emissions. We run these experiments for 10 epochs each and present our results in Table~\ref{table:dpbert1} (averaged over 3 runs). Curiously, the carbon emissions for the $\epsilon=0.5$ case is comparable to the EU's 2021 passenger vehicle standard \cite{noauthor_undated-qp}. The difference between a private model ($\epsilon = 0.5$) and a non-private model ($\epsilon = \infty$) is approximately 1g of CO$_{2}$, which is equivalent to the emissions from five Google searches \cite{googlesearch_carbfootprint}.

\begin{table}[H]
    \small 
    \centering
    \scalebox{0.85}{
    \begin{tabular}{|l|p{0.25\columnwidth}|p{0.25\columnwidth}|l|}
    \hline
    \textbf{Epsilon ($\varepsilon$)} & \textbf{CE (g)} & \textbf{EC (Wh)} & \textbf{Accuracy (\%)} \\ \hline
    0.5 & 26.7 $\pm$ 0.63 & 49.9 $\pm$ 1.2 & 48.5 $\pm$ 1.39 \\ \hline
    2 & 26.3 $\pm$ 0.49 & 49.3 $\pm$ 0.9 & 52.0 $\pm$ 0.73 \\ \hline
    5 & 26.1 $\pm$ 0.1 & 48.9 $\pm$ 0.9 & 52.3 $\pm$ 0.36 \\ \hline
    15 & 25.9 $\pm$ 0.09  & 48.5 $\pm$ 0.1 & 54.2 $\pm$ 1.40\\ \hline
    $\infty$ \textbf{(Non-Private)} & 25.2 $\pm$ 0.00 & 47.1 $\pm$ 0.27 & 58.5 $\pm$ 5.29 \\ \hline
    \end{tabular}
    }
    \caption{\textbf{DP-BERT:} Emission-Accuracy trends over change in $\epsilon$ for reaching 52\% accuracy.}
    \label{table:dpbert1}
\end{table}

In congruence with existing literature, the accuracy of the differentially private BERT increases consistently with the increase in epsilon. Interestingly, with the increase in the epsilon value – both, CE and EC decrease, though not by a very significant margin. Given that the range of the chosen  $\varepsilon$ varies considerably, the consequent difference in the carbon emission is not proportionally varied. The practical implication of this invariance can be seen as incurring nearly the same carbon footprint for two versions of a model with different degrees of privacy.

\begin{table}[H]
    \small
% \newcolumntype{C}[1]{>{\centering\arraybackslash}p{#1}}
    \centering
    \scalebox{0.85}{
  \begin{tabular}{|l|l|l|l|}
    %\toprule
    \hline 
    \textbf{Epsilon ($\varepsilon$)} & \textbf{Epochs} & \textbf{CE (g)} & \textbf{EC (Wh)} \\ \hline %\midrule
    0.5 & 19 & 153.6 & 287.3 \\ \hline
    2 & 12 & 96.6 & 180.6 \\ \hline
    5 & 9 & 80.9 & 151.3 \\ \hline
    15 & 7 & 56.9 & 106.5 \\ \hline
    $\infty$ \textbf{(Non-Private)} & 6 & 8.5 & 16 \\ \hline %\bottomrule
    \end{tabular}
    }
    \caption{Observing the number of epochs needed to achieve the threshold test accuracy ($T$) with different privacy levels}
    \label{table:dpbert2}
\end{table}

\vspace{-5mm}
  
    \subsubsection{Resource Expense Analysis} 
    The main aim of this experiment is to evaluate how many resources, in terms of consequent carbon and energy emissions are expended in order to achieve a target or threshold accuracy with different degrees of privacy. As defined in the previous set of experiments, we compute the accuracies over $\varepsilon=0.5, 2, 5, 15$.We set the target/threshold accuracy ($T$) to $52\%$ as shown in Table~\ref{table:dpbert2}.
    \par It can be inferred from Table \ref{table:dpbert2} that the Carbon Emission and Energy Usage required to attain the maximum experimental value of privacy is nearly 18 times the carbon emission required to attain the same threshold accuracy with a non-privacy preserving variant of the model. The practical consequence of this experiment dictates that enhancing the degree of privacy of the model, can incur a huge compute cost, which can invariably increase the carbon footprint of the model's training pipeline.
     
\begin{figure}[!htb]
    \small 
    \subfigure[\textbf{BERT:} Training Accuracy]{\label{fig:nlp_train}%
    \includegraphics[width=0.9\linewidth,keepaspectratio,scale=0.5]{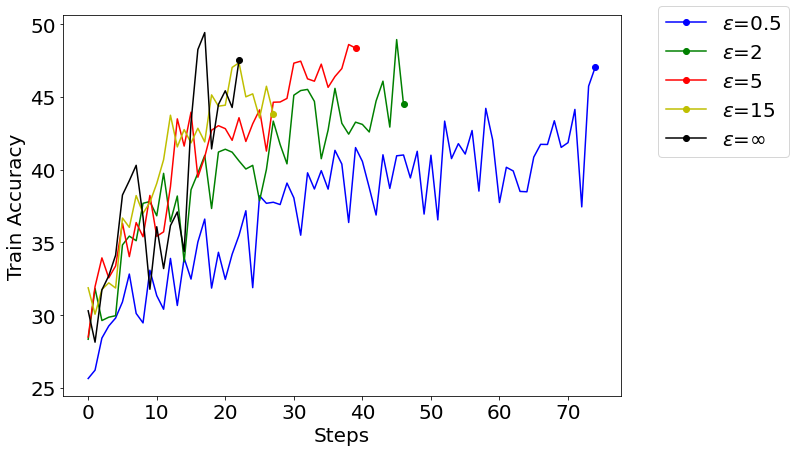}}
    \hfill
    \subfigure[\textbf{BERT:} Testing Accuracy]{\label{fig:nlp_test}%
    \includegraphics[width=0.9\linewidth,keepaspectratio,scale=0.5]{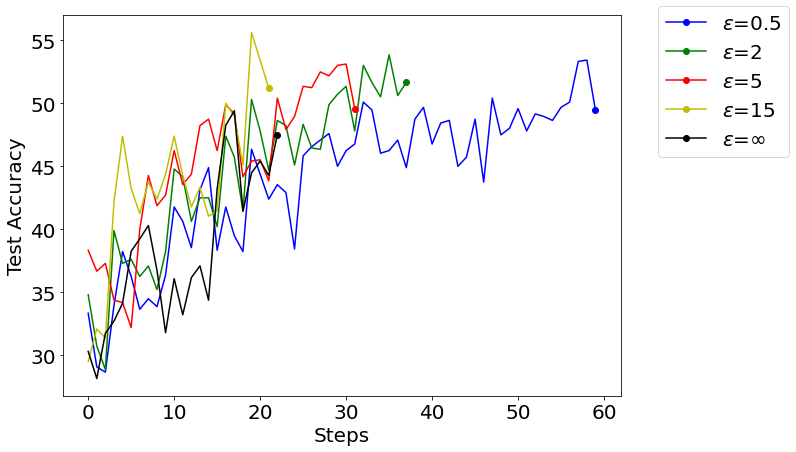}}
   \caption{\textbf{BERT with Gaussian DP:} Training and Testing accuracy trends over change in $\epsilon$ where the threshold accuracy ($T$) is set to 52\%.}
   \label{fig:NLPexpts}
\end{figure}
    
\par Additionally, From Figure~\ref{fig:NLPexpts}, which present the accuracy curves for the experiment, it is quite evident that the vanilla variant (i.e a model without DP-noise) achieves the threshold accuracy with a significantly smaller carbon footprint than all the footprint of its privacy-preserving variants.

% Edited         
% \begin{table}[H]
%     \small
%     \centering
%     \scalebox{0.8}{
%     \begin{tabular}{|l|l|l|l|}
%     \hline
%     \textbf{Epsilon ($\varepsilon$)} & \textbf{Epochs} & \textbf{Carbon Emissions (g)} & \textbf{Energy (Wh)} \\ \hline
%     0.5 & 19 & 153.6 & 287.3 \\ \hline
%     2 & 12 & 96.6 & 180.6 \\ \hline
%     5 & 9 & 80.9 & 151.3 \\ \hline
%     15 & 7 & 56.9 & 106.5 \\ \hline
%     $\infty$ (\textbf{Non-Private)} & 6 & 8.5 & 16 \\ \hline
%     \end{tabular}
%     }
%     \caption{Observing the number of epochs needed to achieve the threshold accuracy ($T$) with different privacy levels}
%     \label{table:dpbert2}
% \end{table}

\subsection{MNIST \& CIFAR}

\begin{table}[!htb]
\small
\centering
% \vspace{1.0ex}
\scalebox{0.85}{
\begin{tabular}{|l|l|l|}
\hline
\textbf{Epsilon ($\varepsilon$)} & \textbf{CE (g)} & \textbf{EC (Wh)} \\ \hline
0.5 \textbf{*} & 10.53 $\pm$ 2.21 & 40.41 $\pm$ 0.93 \\ \hline
2 \textbf{*} & 10.6 $\pm$ 2.43 & 40.5 $\pm$ 0.53 \\ \hline
5 & 7.85 $\pm$ 1.84 & 29.93 $\pm$ 0.46 \\ \hline
15 & 1.61 $\pm$ 0.37 & 6.17 $\pm$ 0.27 \\ \hline
$\infty$ \textbf{(Non-Private)} & 0.08 $\pm$ 7e-04 & 0.38 $\pm$ 3.3e-03 \\ \hline
\end{tabular}
}
\caption{\textbf{MNIST:} Emission trends over change in $\epsilon$ for reaching 70\% accuracy (\textbf{*} \textit{70\% accuracy not reached even after 200 epochs.)}}
\end{table}

\begin{table}[!htb]
\small
\centering
% \vspace{1.0ex}
\scalebox{0.85}{
\begin{tabular}{|l|l|l|}
\hline
\textbf{Epsilon ($\varepsilon$)} & \textbf{CE (g)} & \textbf{EC (Wh)} \\ \hline
0.5* & 15.34 & 70.216 \\ \hline
2 & 12.48 & 57.108 \\ \hline
5 & 3.12 & 14.307 \\ \hline
15 & 2.03 & 9.297 \\ \hline
$\infty$ \textbf{(Non-Private)} & 0.36 & 1.678 \\ \hline
\end{tabular}
}
\caption{\textbf{CIFAR:} Emission trends over change in $\epsilon$ for reaching 55\% accuracy in a single run (\textbf{*} \textit{55\% accuracy not reached even after 30 epochs.)}}
\end{table}

\begin{figure}[!htb]
    \small
    \subfigure[\textbf{MNIST:} Accuracy on Training Set]{\label{fig:mnist_train}%
    \includegraphics[width=0.9\linewidth,keepaspectratio]{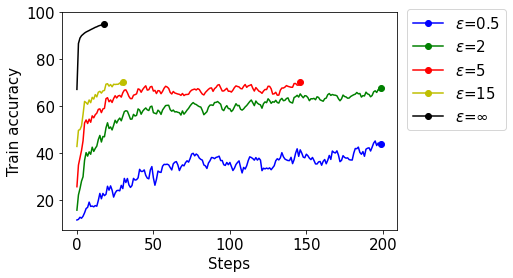}}
    %\hfill
    \subfigure[\textbf{MNIST:} Accuracy on Test Set]{\label{fig:mnist_test}%
    \includegraphics[width=0.9\linewidth,keepaspectratio]{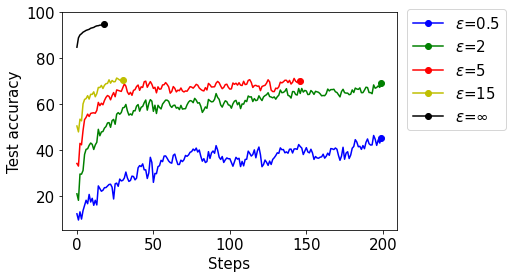}}\vspace{-10pt}
   \caption{\textbf{MNIST with Gaussian DP:} Training and test accuracy trends during training for multiple $\epsilon$ values.}
   \label{fig:MNISTexpts}
   \vspace{-5mm}
\end{figure}

We evaluate our approach on the MNIST dataset \cite{lecun-mnisthandwrittendigit-2010} with a batch size of 128 using DP-SGD \cite{Abadi_2016}. We use a simple multi-layer perceptron (MNIST 2NN) with a two hidden layers of 200 units each (parameters = 199,210) as the network from \cite{mcmahan2017communicationefficient}. Our goal is to observe the trend in the CO$_2$ emissions by allowing the model to train and reach $X$ accuracy with different values of $\varepsilon$ (different levels of privacy).
We compute the accuracies over $\varepsilon=0.5, 2, 5, 15$ as shown in Fig.~\ref{fig:MNISTexpts}.  We set the target/threshold accuracy ($T$) to $70\%$ so that most of the privacy-variant models can achieve under $200$ iterations. In Fig.~\ref{fig:MNISTexpts} we see that only models with $\varepsilon=5, 15$ reach 70\% accuracy within 200 epochs.
Fig.~\ref{fig:MNISTexpts} shows a clear trend on how increasing levels of privacy in ML models increases the amount of computation required to reach $T$, thereby releasing higher carbon emissions. 
% We’ve also compared models trained with privacy and without privacy(epsilon = infinity) and observed a huge increase in the computation requirement and carbon emissions released.

\begin{figure}[!htb]
    \small
    \subfigure[\textbf{CIFAR-10:} Accuracy on Training Set]{\label{fig:cifar_train}%
    \includegraphics[width=0.9\linewidth,keepaspectratio]{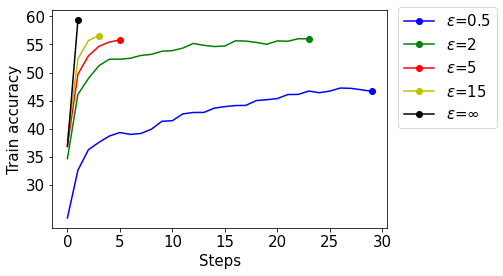}}
    %\hfill
    \subfigure[\textbf{CIFAR-10:} Accuracy on Test Set]{\label{fig:cifar_test}%
    \includegraphics[width=0.9\linewidth,keepaspectratio]{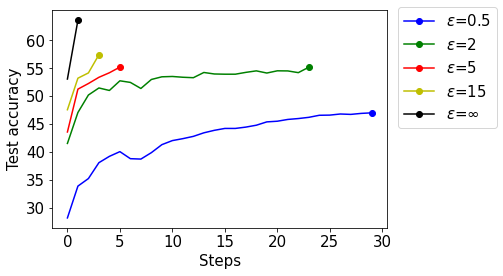}}\vspace{-10pt}
   \caption{\textbf{CIFAR-10 with Gaussian DP:} Training and test accuracy trends during training for multiple $\epsilon$ values.}
   \label{fig:CIFARexpts}
   \vspace{-5mm}
\end{figure}

For the CIFAR-10 experiments, we use the ResNet18 model pre-trained on the ImageNet dataset. This deep variant of CNN is chosen instead of the simple MNIST 2NN in order to capture more complex and realistic features. We compute the accuracies over $\varepsilon=0.5, 2, 5, 15$ as shown in Fig.~\ref{fig:CIFARexpts}.  We set the target/threshold accuracy ($T$) to $55\%$ as most of the private models can achieve under $30$ iterations. We run the CIFAR experiments only for 30 iterations as we observe that there is little to no significant improvement beyond this. In Fig.~\ref{fig:CIFARexpts}, We observe similar trends as Fig.~\ref{fig:MNISTexpts} on how increasing privacy levels involve higher computations to reach $T$ which in turn, release higher carbon emissions. 

%  MNIST Experiments: \newline
%  1) Constant Number of epochs: 50
% \newline
% \begin{table*}[h]
% \centering
% \begin{tabular}{|l|l|l|l|l|}
% \hline
% Epsilon & Train Accuracy & Test Accuracy & Energy Consumption(Wh) & CO2 Emissions (g) \\ \hline
% 2 & 67.57 & 68.29 & 10.71 & 1.49 \\ \hline
% 1 & 60.90 & 62.07 & 11.44 & 1.59 \\ \hline
% 0.5 & 50.14 & 50.73 & 11.11 & 1.54 \\ \hline
% 0.25 & 35.33 & 35.72 & 10.45 & 1.45 \\ \hline
% 0.125 & 19.90 & 20.21 & 11.18 & 1.55 \\ \hline
% \end{tabular}
% \caption{MNIST constant epochs}
% \end{table*}

\subsection{Cartpole}

For the reinforcement learning experiments, we trained a DQN over OpenAI Gym's Cartpole-v0 environment. Due to page restrictions we defer the discussion of the RL experiments to the Appendix B. 

\section{Conclusion}
 We demonstrate and highlight the prominent impact of Privacy-Preserving Machine Learning (PPML) on carbon emissions over three ML domains, namely, CV, NLP and RL. We observe that the stronger privacy regime, \emph{i.e,} a lower $\epsilon$ value, ML algorithms always result in higher levels of carbon emissions in the CV and NLP domains. Curiously, results for RL are less obviously explained and we defer these discussions for future work. We conclude that alongside the challenge of obtaining state-of-the-art performance, PPML needs to reduce the number of epochs required to reach the desired performance. This leads us to the following critical questions which we leave as open questions for the future: (1) Can we reduce the number of iterations (including hyperparameter tuning) required to reach a privacy-utility ratio? (2) How much does the size of ML models affect the carbon emissions and the overall performance under PPML?
 
\section{Acknowledgements}

We thank Fatemehsadat Mireshghallah for her valuable inputs and discussions.

\bibliography{dp}
\bibliographystyle{icml2021}

\appendix
\section{Hyperparameter Tuning}

\subsection{BERT}

We set out to understand the impact of noise introduced by DP on the model convergence up until a threshold test accuracy. We primarily tune the epsilon values (for the privacy guarantee) and the number of epochs. For a fair comparison we use constant hyperparameters for each epsilon so as to to minimize the scope of confounding variables (optimally tuned hyperparameters per epsilon value) for BERT (Table~\ref{table:dpbert1}). 

Succeeding that though, since a lot of differentially private models are deployed in edge-focused setups where numerous clients can take hyperparameter optimization decisions locally - the experiments (Table~\ref{table:dpbert2}) with the thresholding accuracy were conducted. On grounds of ablation, a subset of these locally available hyperparamters were chosen with the objective to evaluate the change in carbon and energy emissions when these hyperparameters are altered: the privacy guarantee (meant to be a globally decided parameter in a distributed setup) and the number of epochs (meant to be a locally decided parameter) are changed. 

\subsection{MNIST \& CIFAR}

Similar to the BERT approach, we use consistent hyperparameters across different $\varepsilon$-privacy levels for the MNIST dataset.

However, due to serious model performance degradation on the CIFAR-10 dataset we use an alternate approach. We tweak a set of hyperparameters to achieve optimal performance at the respective $\varepsilon$-privacy levels when the model performs poorly. For instance, we observe a major drop in performance for only the non-private model when a constant learning rate of $10^{-3}$ is used for both the private and non-private settings. So, we instead choose a learning rate of $10^{-6}$ for the non-private model which leads to optimal performance. 

In our experiments, we also use the RMS-PROP optimizer which leads to more stable results \& faster convergence during training.

\section{Reinforcement Learning: Experiments \& Discussion}

\subsection{Cartpole}

The Cartpole-v0 environment \cite{Cartpole83} consists of an un-actuated joint to a cart. There are two possible actions which involve a force of +1 or -1 being applied to the cart along a friction-less track. The pole starts upright, with the goal of preventing it from falling over. For every time-step that the pole is upright, a reward of +1 is added to the total reward. However, if the pole exceeds 15 degrees from the vertical, or if the cart moves more than 2.4 units from the center, the episode ends. 

\begin{figure}[!htb]
\small
\vskip 0.2in
\begin{center}
\centerline{\includegraphics[width=0.9\linewidth]{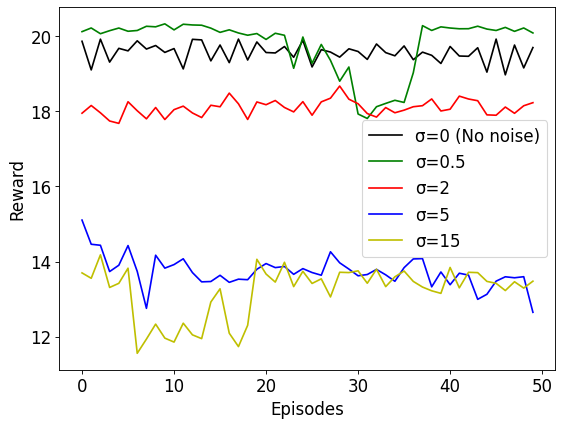}}
\caption{\textbf{CartPole with Gaussian DP:} Episodes vs Rewards for the mean reward every 100 episodes}
\label{fig:gdp_img}
\end{center}
\vskip -0.2in
\end{figure}

\begin{figure}[!htb]
\small
\vskip 0.2in
\begin{center}
\centerline{\includegraphics[width=0.9\linewidth]{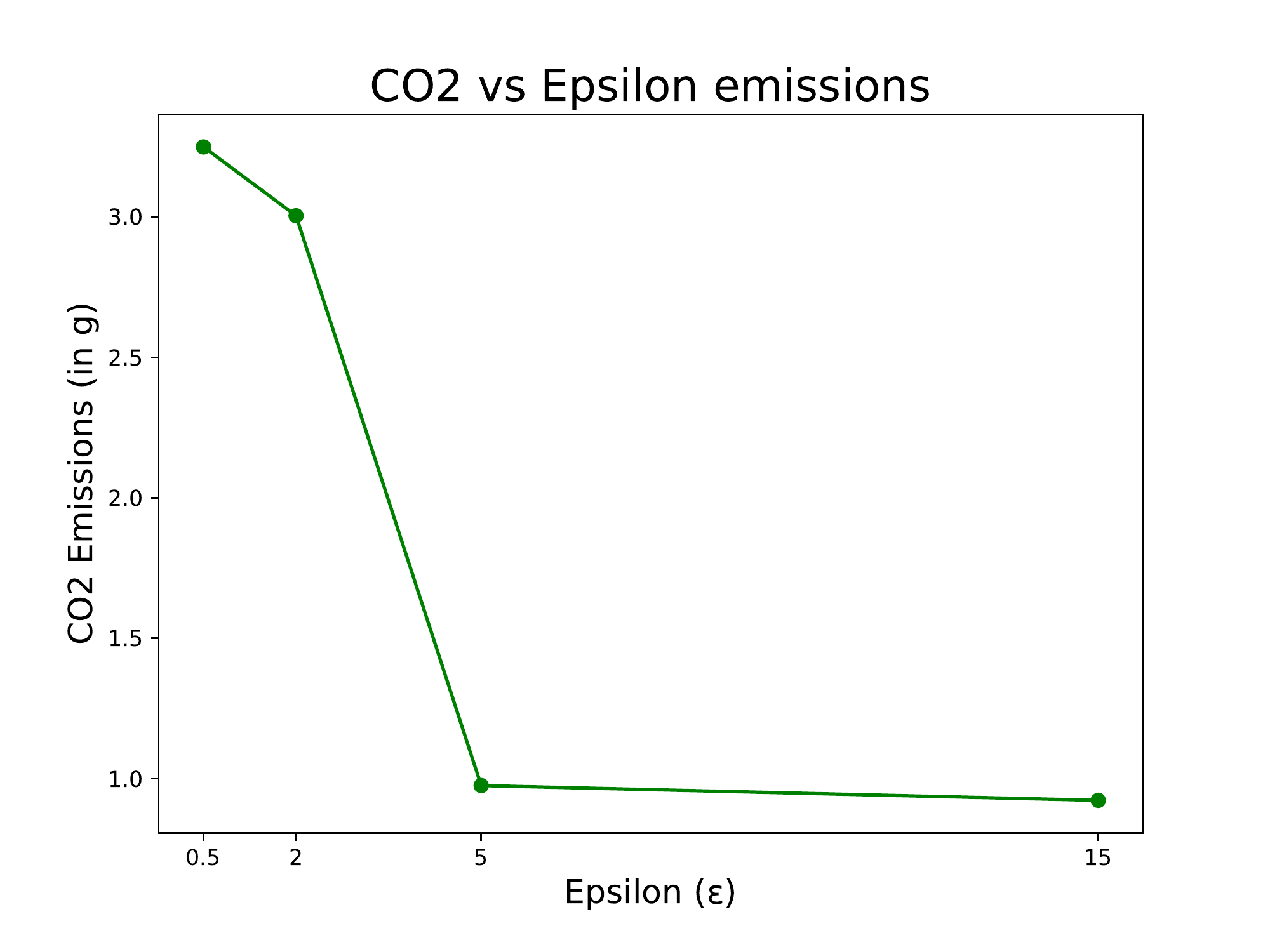}}
\caption{\textbf{Acrobot with Gaussian DP:} CO2 vs Epsilon values post training after a 1000 episodes}
\label{fig:acro_co2_1000}
\end{center}
\vskip -0.2in
\end{figure}

\begin{figure}[!htb]
\small
\vskip 0.2in
\begin{center}
\centerline{\includegraphics[width=0.9\linewidth]{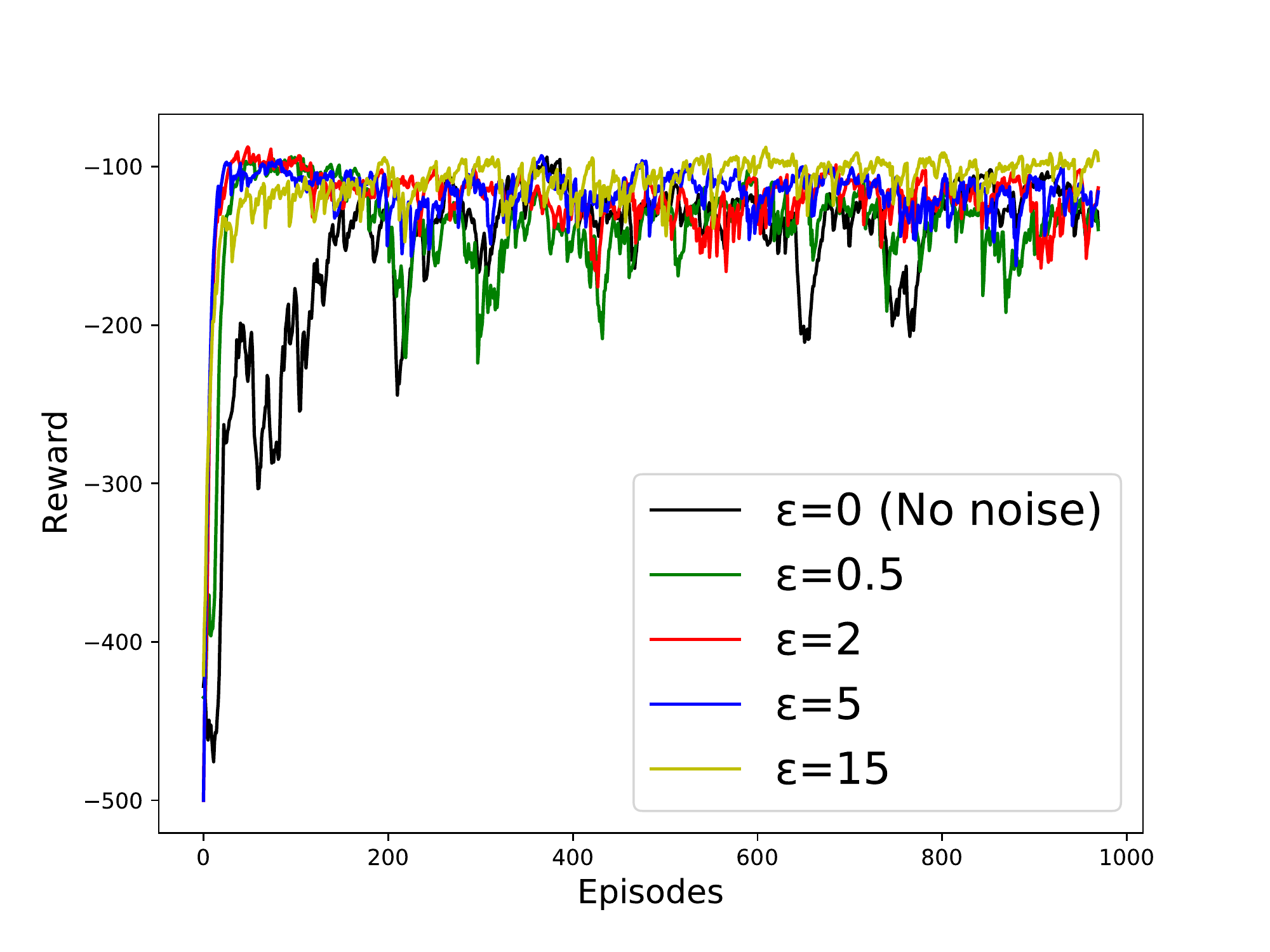}}
\caption{\textbf{Acrobot with Gaussian DP:} Episodes vs Rewards for the mean reward every episode for a 1000 episodes}
\label{fig:acro_avg_1000}
\end{center}
\vskip -0.2in
\end{figure}

The DQN’s configuration (including the hyperparameters) is the same as the one used in \cite{wang2019privacypreserving}, and we observed results similar to this paper, with one variant of DP model slightly outperforming the baseline as shown in \ref{fig:gdp_img}. It consists of a single hidden layer with 16 neurons. For our non-private experiment we obtained a mean reward of 19.94 and carbon emissions of 0.22 g on average (over a 1000 episodes). We provide results of the private variants in Fig.~\ref{fig:gdp_img}. Our setup included multiple experiments.
\begin{itemize}[noitemsep,topsep=0pt]
    \item Noise addition to DQN's output layer only. (1)
    %\item Noise addition to both, the output layer of the DQN as well as the parameters, before gradient clipping was applied
    \item Noise addition to both, the DQN's output layer and its parameters. The noise added to the parameters is the averaged noise sampled from the \textit{noisebuffer} function during the forward pass. (2)
\end{itemize}

We varied the value of the variance {$\sigma$} of the distribution to observe its impact on the function approximated by the DQN. As expected, with increasing noise addition to the model (\emph{i.e.}, increasing value of $\sigma$), we notice a drop in the average reward. Subsequently, the increased computations lead to higher carbon emissions. We observe that there is a significant increase in CE from Table~\ref{gdp1000} to Table~\ref{gdp5000} when the number of episodes increase. 
% \textbf{The addition of functional noise to the model decreases the mean reward and also aids in learning an approximate function in comparison to the baseline experiment \cite{wang2019privacypreserving}.}
% and also prevents the model from learning as approximating the function as well as the one learned by the agent in the baseline experiment.
\subsection{Acrobot}

We alternatively run experiments on the Acrobot-v1 environment. Acrobot-v1 has 2 joints and links and the joint between the links is actuated. The environment’s initial position has the links in a resting state, hanging downward, with the objective being to swing the end of the lowermost link up to a specified height. There are 3 potential actions, to apply positive torque, to apply negative torque, or to do nothing. The agent attempts to maximize its reward within \textit{500} time-steps, with each time-step potentially incurring a punishment of \textit{-1}, thus making \textit{-500} the worst possible reward. If the agent reaches the given height, a reward of \textit{0} is returned. The state returned consists of the joint angular velocities and the sin and cos values of the two rotational joint angles.

For the purpose of training the Acrobot environment, we used distributed reinforcement learning with quantile regression \cite{dabney2017distributional}. Quantile regression models the distribution of the returned rewards instead of only considering the mean of the distribution. We use the Opacus library in PyTorch to add the necessary DP guarantees. From Fig.\ref{fig:acro_co2_1000}, we observe a consistent upward trend of the CO2 emissions with the epsilon value. This directly implies that better privacy guarantees lead to higher carbon emissions. As a sanity check, we can observe (from Table~\ref{AcrobotTable}) that the mean reward increases with higher epsilons (lower privacy requirement) due to reduced noise levels.

\begin{table}[!htb]
\small
\centering
\scalebox{0.85}{
\begin{tabular}{|p{1.5cm}|p{1.5cm}|l|l|}
\hline
$\text{\textbf{Epsilon}}^{*}$ ($\epsilon^{*} \propto 15\epsilon$) & \textbf{Sigma} ($\sigma \propto \frac{1}{\epsilon}$) & \textbf{Mean Reward} & \textbf{CE (g)} \\ \hline
1 & 15               & 4.5 ± 0.6          & 1.03 ± 0.06 \\ \hline
3 & 5                & 2.2 ± 0.2          & 0.96 ± 0.03\\ \hline
7.5 & 2                & 19.9 ± 0.5          & 1.14 ± 0.06  \\ \hline
30 & 0.5              & 19.4 ± 0.1          & 1.15 ± 0.06 \\ \hline
\end{tabular}
}
\caption{\textbf{CartPole:} Emission trends over change in $\epsilon^{*}$ post 1000 episodes in (1) following \cite{wang2019privacypreserving}}
\label{gdp1000parammain}
\end{table}

\begin{table}[!htb]
\small
\centering
\scalebox{0.85}{
\begin{tabular}{|p{1.5cm}|p{1.5cm}|l|l|}
\hline
$\text{\textbf{Epsilon}}^{*}$ ($\epsilon^{*} \propto 15\epsilon$) & \textbf{Sigma} ($\sigma \propto \frac{1}{\epsilon}$) & \textbf{Mean Reward} & \textbf{CE (g)} \\ \hline
1 & 15               & 2.3 ± 0.9            & 0.41 ± 0.01  \\ \hline
3 & 5                & 10.2 ± 0.8          & 0.5 ± 0.02  \\ \hline
7.5 & 2                & 7.6 ± 0.7           & 0.45 ± 0.02 \\ \hline
30 & 0.5              & 13.8 ± 0.1          & 0.48 ± 0.03 \\ \hline
\end{tabular}}
\caption{\textbf{CartPole:} Emission trends post 1000 episodes in (2)}
\label{gdp1000}
\end{table}

\begin{table}[!htb]
\small
\centering
\scalebox{0.8}{
\begin{tabular}{|p{1.5cm}|p{1.5cm}|l|l|}
\hline
$\text{\textbf{Epsilon}}^{*}$ ($\epsilon^{*} \propto 15\epsilon$) & \textbf{Sigma} ($\sigma \propto \frac{1}{\epsilon}$) & \textbf{Mean Reward} & \textbf{CE (g)}  \\ \hline
1 & 15               & 13.2 ± 0.3          & 3.51 ± 0.26  \\ \hline
3 & 5                & 13.7 ± 0.9          & 2.31 ± 0.28  \\ \hline
7.5 & 2                & 18.1 ± 0.1           & 2.72 ± 0.23  \\ \hline
30 & 0.5              & 19.8 ± 0.6          & 4.0 ± 0.31 \\ \hline
\end{tabular}}
\caption{\textbf{CartPole:} Emission trends post 5000 episodes in (2)}
\label{gdp5000}
\end{table}

\begin{table}[!htb]
\small
\centering
\scalebox{0.85}{
\begin{tabular}{|l|l|l|}
\hline
\textbf{Epsilon} & \textbf{Mean Reward} & \textbf{CO2} \\ \hline
0.5              & -150.7 ± 18.69       & 3.25 ± 0.372 \\ \hline
2                & -138.0 ± 11.15       & 3.0 ± 0.245  \\ \hline
5                & -120.6 ± 4.64        & 0.98 ± 0.035 \\ \hline
15               & -117.5 ± 2.45        & 0.92 ± 0.023 \\ \hline
0                & -110.5 ± 1.22        & 0.33 ± 0.023 \\ \hline
\end{tabular}
}
\caption{\textbf{Acrobot-v1}: Emission trends post a 1000 Episodes}
\label{AcrobotTable}
\end{table}

%\begin{table}[!htb]
%\small 
%\centering
%\scalebox{0.8}{
%\begin{tabular}{|p{1.5cm}|p{1.5cm}|l|l|}
%\hline
%$\text{\textbf{Epsilon}}^{*}$ ($\epsilon^{*} \propto 3\epsilon$) & \textbf{Sigma} ($\sigma \propto \frac{1}{\epsilon}$) & \textbf{Mean Reward} & \textbf{CO2 (g)} \\ \hline
%1 & 15               & 1.3 ± 0.04           & 0.84 ± 0.018 \\ \hline
%3 & 5                & 4.5 ± 1.18           & 0.97 ± 0.083 \\ \hline
%7.5 & 2                & 10.6 ± 0.15          & 1.05 ± 0.061 \\ \hline
%30 & 0.5              & 4.8 ± 2.2            & 0.88 ± 0.004 \\ \hline
%\end{tabular}}
%\caption{GDP with noise added to parameters before gradient clipping}
%\label{gdpparams1}
%\end{table}

\end{document}